\documentclass[12pt]{iopart}
\usepackage{graphicx,amssymb,dsfont,bm}
\usepackage{iopams} 
\usepackage{color}
\usepackage{multirow}
\usepackage{makecell}
\usepackage{enumitem}
\usepackage{citesort}
\usepackage{longtable}
\usepackage{changepage}
\usepackage{tikz}
\usetikzlibrary{arrows.meta}
\usetikzlibrary{shadows}
\usepackage{colortbl,dcolumn}
\usepackage{tikz}
\usepackage{collcell}
\usepackage{xcolor}
\usepackage{pgf}
\usepackage{collcell}
\usepackage{boldline}
\usepackage{longtable}

\usepackage[titletoc,page]{appendix}
\let\appendixpagenameorig\appendixpagename
\renewcommand{\appendixpagename}{\normalsize\appendixpagenameorig}

\usepackage[titles]{tocloft}
\def\eqref#1{(\ref{#1})}

\begin{document}

\title[Grasping asymmetric information in market impacts]{Grasping asymmetric information in market impacts}
\author{Shanshan Wang$^1$, Sebastian Neus\"u\ss$^2$ and Thomas Guhr$^1$}
\address{$^1$Fakult\"at f\"ur Physik, Universit\"at Duisburg--Essen, Lotharstra\ss e 1, 47048 Duisburg, Germany}
\address{$^2$Deutsche B\"orse AG, Frankfurt, Germany}
\ead{shanshan.wang@uni-due.de}
\vspace{10pt}
\begin{indented}
\item[]\today
\end{indented}

\begin{abstract}
The price impact for a single trade is estimated by the immediate response on an event time scale, \textit{i.e.}, the immediate change of midpoint prices before and after a trade. We work out the price impacts across a correlated financial market. We quantify the asymmetries of the distributions and of the market structures of cross-impacts, and find that the impacts across the market are asymmetric and non-random. Using spectral statistics and Shannon entropy, we visualize the asymmetric information in price impacts. Also, we introduce an entropy of impacts to estimate the randomness between stocks. We show that the useful information is encoded in the impacts corresponding to small entropy. The stocks with large number of trades are more likely to impact others, while the less traded stocks have higher probability to be impacted by others.
\end{abstract}

\begin{adjustwidth}{2.5cm}{}
\end{adjustwidth}

\maketitle

\pagestyle{empty}
\noindent\rule{\textwidth}{1pt}
\vspace*{-1cm}
\tableofcontents
\noindent\rule{\textwidth}{1pt}
\pagestyle{headings}

\section{Introduction}
\label{sec1}

In the last two decades, the microstructure of financial markets has attracted ever more attention. An enormous amount of available transaction data makes quantitative analyses possible. Since the 1960s when Mandelbrot found the fat-tailed distribution of cotton prices~\cite{Mandelbrot1963}, many stylized facts in the price dynamics~\cite{Cont2001,Schmitt2012,Gabaix2003} were identified. In particular, the price change caused by trades exhibits non-Markovian features~\cite{Bouchaud2004,Lillo2003,Wang2016a,Wang2016b}, which drive the market to temporarily deviate from an efficient state~\cite{Wang2016a}. The average price change due to a trade is referred to as price impact~\cite{Bouchaud2010}. The price impact in single stocks is more likely to induce an extra cost of trading, termed liquidity cost~\cite{Demsetz1968}. To reduce such costs, traders split orders, which partially leads to the long-memory correlation in the order flow~\cite{Lillo2005,Bouchaud2004}. The impact from a large split order is termed market impact~\cite{Almgren2005,Torre1997}. These findings have considerable practical and theoretical importance. 

Recently, empirical studies~\cite{Wang2016a,Wang2016b,Benzaquen2017} disclosed that there are also price impacts across stocks. To avoid confusion, the impact in single stocks is named self-impact and the impact between stocks is named cross-impact. Different from the costs arising from the self-impact, the extra costs \cite{Schneider2018,Wang2017,Mastromatteo2017} caused by the cross-impact are often ignored in the optimal execution of orders~\cite{Gatheral2010,Gatheral2012,Gatheral2013,Obizhaeva2013,Alfonsi2014,Alfonsi2016}. From a ``no-dynamic-arbitrage'' perspective, the cross-impact should be symmetric~\cite{Schneider2018}. However, the empirical cross-impact from asset $i$ to asset $j$ is not equal to the one from asset $j$ to asset $i$~\cite{Schneider2018}. Regardless of the bid-ask spread, the asymmetry of cross-impacts implies that, first, the information distributed in the market is asymmetric, and second, arbitrage is possible if using special strategies. 

The self-impact has been more extensively studied and estimated than the cross-impact, partially due to the difficulties in the empirical estimation. Without a proper estimation, biases may be present either in the cross-impact or in the costs arising from it. We empirically analyzed the cross-response and the cross-impact on a physical time scale~\cite{Wang2016a,Wang2016b,Wang2016c} with a one-second resolution. In their study, Benzaquen \textit{et al.} used time intervals of five minutes~\cite{Benzaquen2017}. Both choices of time scales have advantages and limitations. Schneider \textit{et al.} therefore use the combined trade time~\cite{Schneider2018}. In the present study, we focus on the immediate responses. More precisely, we analyze how a subsequent quote change in one stock is caused by a trade in another stock. In this sense, we use an event time scale. The immediate response helps to conveniently estimate the impact without a time lag. This impact leads to an increase in transaction costs immediately and obviously compared to the impact with a time lag. If we were able to grasp the important information of the immediate impact, we could use this information in trading strategies of multiple stocks for reducing transaction costs, or in risk management for setting an alert line. We analyze these impacts across the whole, correlated market and identify and quantify the asymmetry of the information. Furthermore, the Shannon entropy~\cite{Shannon1948} helps us to assess the degree of randomness for the impacts and to extract other useful features.

The paper is organized as follows. In Sect.~\ref{sec2}, we discuss the data. In Sect.~\ref{sec3}, we measure the asymmetry of impacts across the market. In Sect.~\ref{sec4}, we analyze the spectral statistics of asymmetric impacts. To estimate the degree of randomness, we introduce the Shannon entropy and construct directional networks of impacts with given entropy matrices in Sect.~\ref{sec5}. The conclusions are presented in Sect.~\ref{sec6}.

\section{Data description}
\label{sec2}

We introduce the data set in Sect.~\ref{sec2.1}, and explain the order reconstruction from the historical data in Sect.~\ref{sec2.2}. We then describe the procedure of data processing in Sect.~\ref{sec2.3}.

\subsection{Data set}
\label{sec2.1}

We use the TotalView-ITCH data set, where 96 stocks from NASDAQ stock market, in NASDAQ 100 index are listed. The TotalView-ITCH data set contains the order flow data with all the events, for instance, the submissions, cancellations and executions of limit orders. It has a  resolution of one millisecond that is much higher than the resolution of one second in the Trade and Quote (TAQ) data set~\cite{TAQ2008}. Hence, plenty of order flow data in each trading day are recorded in the TotalView-ITCH data set. For each stock, we take into account the intraday data of five trading days from March 7th to March 11th of 2016, obtained from Tradingphysics~\cite{TP}. Unlike the TAQ data set, the TotalView-ITCH data set does not provide the information of quotes and trades directly. To gain these information, reconstruction of the order book is required. In addition, the trade and quote data used in our study are restricted to the intraday trading time from 9:40 to 15:50 EST. The stocks and the average daily number of trades during this period are listed in~\ref{appA}.

\subsection{Order reconstruction}
\label{sec2.2}

The basic idea for the order reconstruction is to simulate the trading of the stock market with the historical order flow data while organizing the orders into the order book. The historical order flow data can be found in the TotalView-ITCH data set, which records all limit orders with the message for processing. The limit orders enter the order book time-ordered and wait for a better trade price. They are distinguished from another type of orders, \textit{i.e.}, the market orders, which are executed immediately at the present available price.

To begin with, we download the order flow data from the TotalView-ITCH data set and inject the limit orders into an order pool. The order pool is a place where all the limit orders are gathered and processed according to the message carried by each order. The message includes eight types, submission to buy (B), submission to sell (S), cancellation in part (C), cancellation in full (D), execution in part (E), execution in full (F), bulk volumes for cross events (X) and executions of non-display orders (T). We ignore the types X and T, because of the difficulty to identify the trade types and because they are sparse compared to the other types. In the order pool, a submitted limit order is placed at a price level by the principle of primary price priority and secondary time priority. The submitted order will raise the available volume at that price level. However, if a cancellation or an execution is released to an order, the volume at the price level at which the order is placed is reduced in part or in full. To trace an order with different messages, we follow the unique ID given to an order when it enters the order pool. Orders with smaller ID numbers are submitted earlier. If the volume at a certain price level vanishes completely, the order ID and the corresponding price are deleted from the order pool. Therefore,  a message issued to an order will change either the price or the volume. To make such information visible, all prices and corresponding volumes are listed in the order book. It is updated to a new arrangement by a new message, such that the minimal (maximal) price to sell (buy), \textit{i.e.}, the best ask (bid), with the lowest ID number is always listed at the beginning of the ask (bid).

We say there is a new best quote if either the best ask price, the best bid price, the best ask volume or the best bid volume is changed. In this way, we are able to filter the best quote data. We also notice that an execution of a limit order matches a trade of a market order. The trade type of the limit order is opposite to the one of the market order, but the trade price as well as the traded volume for the two orders coincide. Using the message of limit orders, we derive the trade information and obtain the trade data. The data can be distinguished on the level of one millisecond. To facilitate the data processing in the following, we only consider events with a single trade in a given millisecond. If there are more than one trade in one-millisecond interval, we exclude all these trades in that interval. The average proportion of excluded trades to total trades is $10.44\%$.

\subsection{Data processing}
\label{sec2.3}

\begin{figure*}[tbp]
  \begin{center}
    \includegraphics[width=0.75\textwidth]{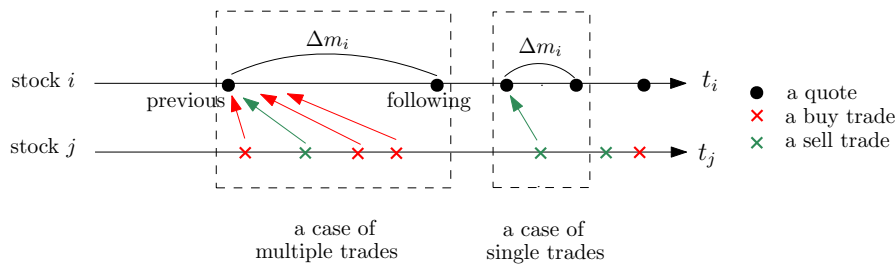}
     \caption{Examples for the cases of multiple trades and single trades, $t_i$ and $t_j$ are the event times of stocks $i$ and $j$, respectively, $\Delta m_i$ is the price change between the previous and the following quotes of stock $i$ for a given trade of stock $j$.}
   \label{fig2.2}
   \end{center}
\end{figure*}

To estimate the immediate change in the quotes due to a trade, we track each trade. If a trade of stock $j$ occurs at time $t$ (measured on the event scale), the last quote of stock $i$ in time $t-1$ is treated as the previous quote of that trade and accordingly the first quote of stock $i$ in time $t+1$ as the following quote. If the latter quote is generated while a trade of stock $i$ occurring, this trade is regarded as being triggered by the preceding trade of stock $j$ at time $t$. Thus, the quote change of stock $i$ can be attributed to the trade of stock $j$. 

In fact, the previous and following quotes are not always directly before and after a trade. There may be several time stamps before or after the trade, such that multiple trades of stock $j$ may share the same previous quote as well as the following quote. We name them cases of multiple trades, shown in Fig.~\ref{fig2.2}. Hence, a single trade is not enough to cause the price change of another stock until several trades occur. Still, there is a part of trades that does not share the previous and following quotes with others. We name them cases of single trades, see Fig.~\ref{fig2.2}. We estimate the average probabilities for the trades belonging to the cases of multiple trades and of single trades as 0.35 and 0.65, respectively. Despite the lower proportion compared to the case of single trades, the case of multiple trades contains a part of the daily transactions. Thus, we take both cases into account.

\section{Asymmetry of price impacts}
\label{sec3}

We employ a price response function to measure the impacts across stocks and to distinguish four types of responses in Sect.~\ref{sec3.1}. For the whole market, we measure the asymmetry of distributions of responses in Sect.~\ref{sec3.2} and also quantify the structural asymmetry of response matrices in Sect.~\ref{sec3.3}.

\subsection{Measurement of impacts}
\label{sec3.1}

The response function in Refs.~\cite{Wang2016a,Wang2016b} measures how a buy or sell order at time $t$ influences on average the price at a later time $t+\tau$. The physical time scale was chosen since the trades in different stocks are not synchronous. Here, we rather use a response function on an event time scale, as we are interested in the immediate responses. The time lag $\tau$ is restricted to one such that the price response quantifies the price impact of a single trade. We define it as 
\begin{equation}
R_{ij}=\Big\langle \Big(\log m_{i}^{(f)}(t_j)-\log m_{i}^{(p)}(t_j)\Big)\varepsilon_j(t_j) \Big\rangle_{t_j}  \ 
\label{eq3.1.1}
\end{equation}
for the price change of stock $i$ caused by a trade of stock $j$. Here, $m_{i}^{(p)}(t_j)$ is the midpoint price of stock $i$ previous to the trade of stock $j$ at its event time $t_j$ and $m_{i}^{(f)}(t_j)$ is the midpoint price of stock $i$ following that trade. The trade sign $\varepsilon_j(t_j)$ of stock $j$ is defined either as $+1$ for a buy or as $-1$ for a sell market order. On the event time scale, zero trade signs are absent. The sign of each trade can be obtained empirically from TotalView-ITCH data set. The symbol $\langle \cdots \rangle_{t_j}$ indicates an average over the event time $t_j$. 

It is worthwhile relating the response~\eqref{eq3.1.1} to the model set up in Ref.~\cite{Wang2016c}. The response $R_{ij}(\tau)$ for a stock pair ($i$, $j$) with a time lag $\tau$ is modelled by 
\begin{equation}
R_{ij}(\tau) = R_{ij}^{(s)}(\tau) + R_{ij}^{(c)}(\tau) \ ,
\label{eq3.1.2}
\end{equation}
where $R_{ij}^{(s)}(\tau) $ is related to self-impacts $G_{ii}(\cdot)$ and cross-correlators of trade signs $\Theta_{ij}(\cdot)$ and $\Theta_{ji}(\cdot)$
\begin{eqnarray} \nonumber
R_{ij}^{(s)}(\tau)&=& \sum_{t\leq t'<t+\tau} G_{ii}(t+\tau-t')\left\langle f_i\big(v_i(t')\big)\right\rangle_t \Theta_{ij}(t'-t)\\  \nonumber
		 &+&~ \sum_{t'<t}\Big[ G_{ii}(t+\tau-t')-G_{ii}(t-t')\Big]\left\langle f_i\big(v_i(t')\big)\right\rangle_t \\
		 &\cdot&\Theta_{ji}(t-t') \ ,
\label{eq3.1.3}
\end{eqnarray}
and $R_{ij}^{(c)}(\tau)$ is related to cross-impacts $G_{ij}(\cdot)$ and self-correlators of trade signs $\Theta_{jj}(\cdot)$
\begin{eqnarray} \nonumber
R_{ij}^{(c)}(\tau)&=&\sum_{t\leq t'<t+\tau} G_{ij}(t+\tau-t')\left\langle g_i\big(v_j(t')\big)\right\rangle_t \Theta_{jj}(t'-t)\\  \nonumber
		&+&~\sum_{t'<t}\Big[ G_{ij}(t+\tau-t')-G_{ij}(t-t')\Big]\left\langle g_i\big(v_j(t')\big)\right\rangle_t \\
		&\cdot& \Theta_{jj}(t-t') \ .
\label{eq3.1.4}
\end{eqnarray}
Here, $\langle f_i(v_i(t'))\rangle_t $ and $\langle g_i(v_j(t'))\rangle_t$ are the impacts of traded volumes of stocks $i$ and $j$ on stock $i$, respectively. With respect to the price impact for a single trade, $\tau=1$. This leads to 
\begin{eqnarray}\nonumber
R_{ij}(1)&=& G_{ii}(1)\left\langle f_i\big(v_i(t)\big)\right\rangle_t \Theta_{ij}(0) \\
		&+&G_{ij}(1)\left\langle g_i\big(v_j(t)\big)\right\rangle_t \Theta_{jj}(0) \ .
\label{eq3.1.5}
\end{eqnarray}
Let $\mathcal{G}_{ij}$ be the price impact of a single trade between stocks $i$ and $j$, which contains the effects due to the trade volumes of stock $j$, and let $\Theta_{ij} $ be the correlator of trade signs without any time lag. Equation~\eqref{eq3.1.5} then is transformed into
 \begin{equation}
R_{ij}=\mathcal{G}_{ii}\Theta_{ij}+\mathcal{G}_{ij}\Theta_{jj} =\sum_{n=i,j}\mathcal{G}_{in}\Theta_{nj} \ .
\label{eq3.1.6}
\end{equation}
Here, only the stocks $i$ and $j$ ($i\neq j$) are considered. For the whole market, the price change of stock $i$ can be regarded as the result of price impacts from all stocks $n$, $n=1,\cdots, N$.  Consequently, the price response for a single trade between stocks $i$ and $j$ is expressed as
\begin{equation}
R_{ij}=\sum_{n=1}^{N}\mathcal{G}_{in}\Theta_{nj} \ .
\label{eq3.1.7}
\end{equation}
Let $\mathcal{G}$ and $\Theta$ be the $N\times N$ impact and correlator matrices with entries $\mathcal{G}_{in}$ and $\Theta_{nj} $, respectively. The immediate responses for the whole market then can be formulated as a matrix product, where the $N\times N$ response matrix $R$ with entries $R_{ij}$ reads,
\begin{equation}
R=\mathcal{G}\Theta \ .
\label{eq3.1.8}
\end{equation}
We mention that for $i=j$ in Eq.~\eqref{eq3.1.7}, self-impacts prevail over cross-impacts, so that $R_{ii}$ approximates to $\mathcal{G}_{ii}\Theta_{ii}=\mathcal{G}_{ii}$ where $\Theta_{ii}=1$. For $i\neq j$, Eq.~\eqref{eq3.1.7} approximates to $R_{ij}=\mathcal{G}_{ii}\Theta_{ij}+\mathcal{G}_{ij}$, as the cross-terms without the self-impact or the sign self-correlator are too trivial. In our study, the self-impact $\mathcal{G}_{ii}$ from a trade of stock $i$ might occur at the moment of updating the following quote or might be absent between the previous and the following quotes. If we treat this trade as being triggered by the preceding trade of a different stock $j$, the potential self-impact is incorporated into the cross-impact and the term $\mathcal{G}_{ii}\Theta_{ij}$ vanishes. From this perspective, the price cross-response $R_{ij}$ directly measures the cross-impact $\mathcal{G}_{ij}$.

\begin{figure*}[tbp]
  \begin{center}
    \includegraphics[width=0.95\textwidth]{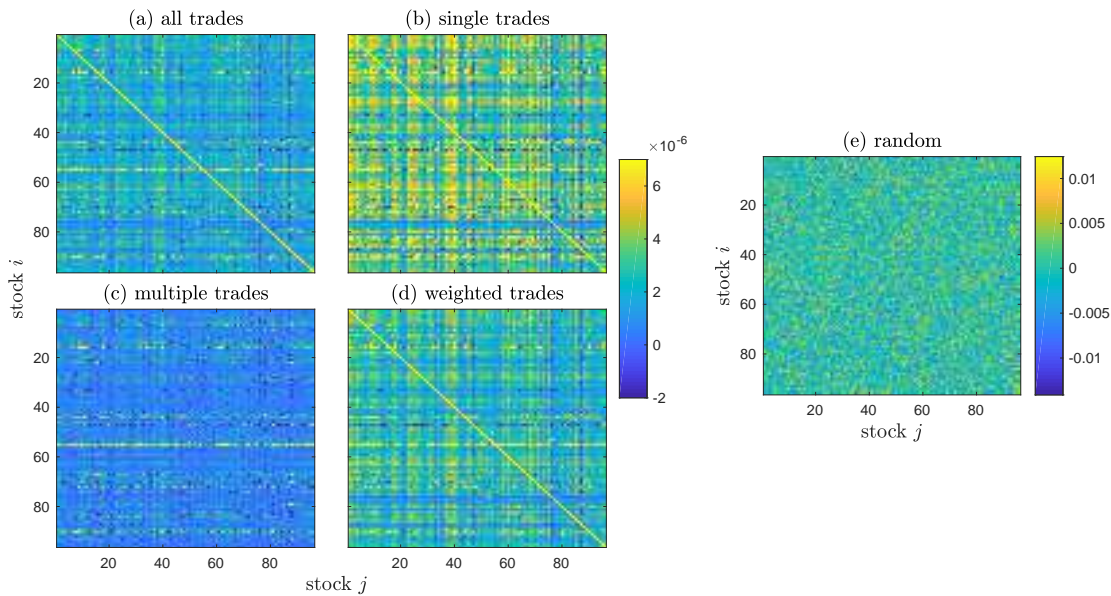}
     \caption{Market response matrices $R$ in (a) case of all trades, (b) case of single trades, (c) case of multiple trades, (d) case of weighted trades, and (e) random case. }
   \label{fig3.1}
   \end{center}
\end{figure*}

By performing different averages, we distinguish the price responses to all trades $R_{ij}|_\mathrm{at}$, to single trades $R_{ij}|_\mathrm{st}$ and to multiple trades $R_{ij}|_\mathrm{mt}$, respectively. Since every trade in this study is classified either as the case of single trades or as the case of multiple trades, the averaging over all trades treats the two cases on equal footing. However, the occurrence of the two cases is never exactly the same for every stock pair. Let $w_{ij}$ be the ratio of the case of single trades to all trades for a stock pair ($i$, $j$). We define a linearly interpolating weighted price response with $w_{ij}$ as a weight factor
\begin{equation}
R_{ij}\Big |_\mathrm{wt}=w_{ij}R_{ij}\Big |_\mathrm{st} + (1-w_{ij}) R_{ij} \Big |_\mathrm{mt} \ .
\label{eq3.1.9}
\end{equation} 
Thus, the frequent occurrence of either case will largely affect the weighted response. $R$ is the $N\times N$ response matrix for all pairs of stocks ($i$, $j$), where in our case $N=96$. In the response matrix, the diagonal elements are the self-responses, and the off-diagonal elements are the cross-responses. We work out the empirical response matrices $R$ for the cases of all trades, single trades, multiple trades and weighted trades, shown in Fig.~\ref{fig3.1}. As seen, the market responds strongly to the case of single trades but much weaker to the case of multiple trades. In between are the case of all trades and the weighted case.

\begin{figure*}[ptb]
\begin{center}
    \includegraphics[width=0.85\textwidth]{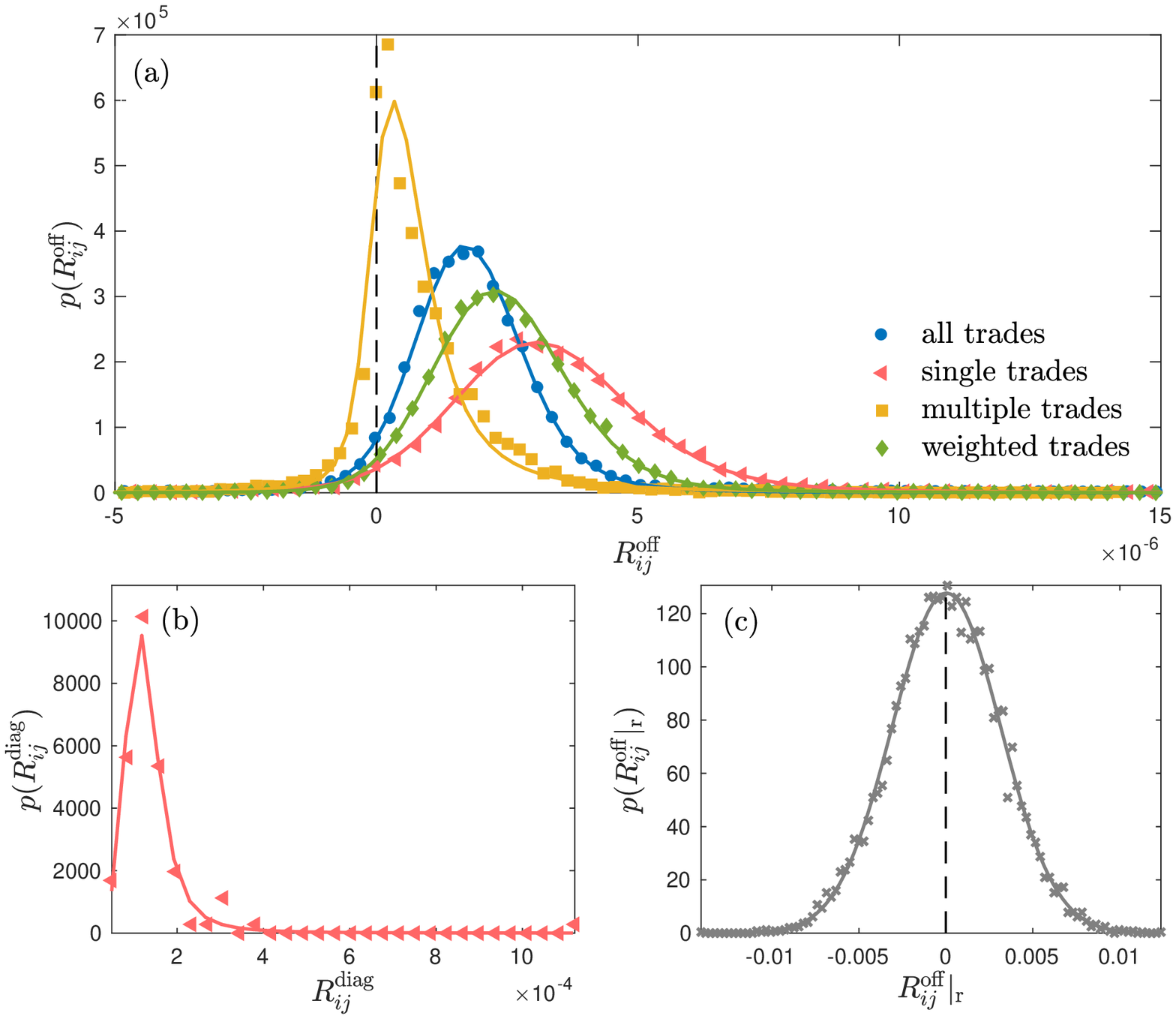}
   \caption{(a) The probability distributions $p(R_{ij}^{\mathrm{off}})$ of price cross-responses to all, to single, to multiple and to weighted trades, respectively; (b) the probability distribution $p(R_{ij}^{\mathrm{diag}})$ of price self-responses to single trades; (c) the probability distribution $p(R_{ij}^{\mathrm{off}}|_\mathrm{r})$ of the off-diagonal elements in the random response matrix $R|_\mathrm{r}$. All distributions are fitted by stable distributions, shown with solid lines.}
   \label{fig3.2}
   \end{center}
\end{figure*}

\begin{table*}[htbp]
\newcolumntype{C}{>{\centering\arraybackslash}p{0.095\textwidth}}
\newcolumntype{L}{>{\arraybackslash}p{0.25\textwidth}}
\begin{center}
\caption{Measurements of asymmetries}
\begin{footnotesize}
\renewcommand\arraystretch{1}
\begin{tabular*}{\textwidth}{CLCCCCCC}
\hlineB{2}
Quantities &Measurements & \makecell{All\\ trades} & \makecell{Single\\ trades} & \makecell{Multiple\\ trades} & \makecell{Weighted\\ trades}& Random\\
\hline
\multirow{4}{*}{\makecell{~~~$p(R_{ij}^{\mathrm{off}})$}}
						 	&Mode/right shift ($\times 10^{-6}$)	&1.965 	&2.710 	& 0.217 	&2.234 & 84.500\\
 							&Mean ($\times 10^{-6}$)	&1.873 &3.348 & 0.793 & 2.473 & 26.923 \\
 							&Median ($\times 10^{-6}$) & 1.775 & 3.168 & 0.508 & 2.354 & 21.304\\
							&Skewness & 1.585 & 1.461	& 1.893	& 1.587 & -0.018\\
\makecell{$\langle\Lambda (R)\rangle$} &Overall asymmetry  & 0.360	& 0.334	& 0.628	& 0.317 & 0.716\\
$H(\mathrm{Im}(\lambda))$ & Shannon entropy   & 2.018 & 1.884 & 2.120 & 1.816 & 3.189\\
\hlineB{2}
\end{tabular*}
\end{footnotesize}
\label{tab3.1}
\end{center}
\end{table*}

\begin{table*}[htbp]
\newcolumntype{C}{>{\centering\arraybackslash}p{0.13\textwidth}}
\newcolumntype{L}{>{\arraybackslash}p{0.17\textwidth}}
\begin{center}
\caption{Fit parameters of stable distributions}
\begin{footnotesize}
\renewcommand\arraystretch{1}
\begin{tabular*}{\textwidth}{CLCCCC}
\hlineB{2}
Responses		&Cases 	& $\alpha$ & $\beta$ & $\gamma$ ($\times 10^{-6}$) & $\mu_0$ ($\times 10^{-6}$) \\
\hline
\multirow{4}{*}{Cross-responses}
			&All trades 		& 1.749 	& 0.512 	&0.748 		& 1.725 \\
			&Single trades 		& 1.737 	& 0.496 	& 1.230 		& 3.107 \\
			&Multiple trades 	& 1.246 	& 0.663 	& 0.480 		& 0.425 \\
			&Weighted trades 	& 1.792	& 0.607 	& 0.915 		& 2.298\\
Self-responses			&Single trades		& 1.493	& 1		& 28.798		& 116.119 \\
Cross-responses		&Random			& 1.999 	& -1		& 2208.170 	& 31.333	\\
\hlineB{2}
\end{tabular*}
\end{footnotesize}
\label{tab3.2}
\end{center}
\end{table*}

For comparison, we also consider a random response matrix $R|_\mathrm{r}$,
\begin{equation}
R|_\mathrm{r}=\frac{1}{L}A~\mathrm{sgn}(B^T) \ ,
\label{eq3.1.10}
\end{equation}
where $A$ and $B$ are uncorrelated $N\times L$ random matrices with zero mean and unit variance, $L$ is the length of a time series. The sign function $\mathrm{sgn}(\cdot)$ is used to obtain random signs from a series of random numbers, hence, $\mathrm{sgn}(B^T)$ is the $L\times N$ matrix of the signs of $B^T$. The superscript $T$ indicates the adjoint. Different from the other cases, the random response matrix $R|_\mathrm{r}$ in Fig.~\ref{fig3.1} (e) displays a uniform distribution without any striking feature.

\subsection{Asymmetry of distributions}
\label{sec3.2}

To quantify the response structure of the whole market, we work out the probability distributions for the four types of responses, see Fig.~\ref{fig3.2}. The cross-responses are the off-diagonal elements of the response matrix $R$ and the self-responses are the diagonal elements. The distribution for the off-diagonal elements of the random response matrix $R|_\mathrm{r}$ is shown in Fig.~\ref{fig3.2} as well. The modes, means, medians and skewness for the distributions of cross-responses are listed in Table~\ref{tab3.1}. Each non-random empirical distribution is shifted to the right of the vertical axis at zero. This asymmetry reveals an imbalance of positive and negative responses. It implies that a buy (sell) of one stock is more likely to move up (down) the price of another stock. Although the cross-response to single trades has the largest right shift of $2.71\times 10^{-6}$ among the four types, it is very weak compared to the self-response with a shift of $1.23\times 10^{-4}$. 

We fit all empirical distributions with stable distributions, a class of probability distributions modelling skewness and heavy tails~\cite{Nolan2003}. Stable distributions $p(x)$ of a random variable $x$ are best specified by their characteristic function 
\begin{equation}
\varphi(\kappa)=\int\limits_{-\infty}^{+\infty} \exp(i\kappa x) p(x) dx \ ,
\label{eq3.2.1}
\end{equation}
which have the form~\cite{Nolan2003} 
{\setlength{\mathindent}{0cm}
\begin{equation}
\varphi(\kappa)=\left\{
\begin{array}{ll}
\exp\left(-\gamma^{\alpha}|\kappa |^{\alpha}\Big[1+i\beta \mathrm{sgn}(\kappa )\tan\frac{\pi \alpha}{2}\big((\gamma|\kappa |)^{1-\alpha}-1\big)\Big] + i\mu_0\kappa \right) & \mathrm{for} \quad \alpha\neq1 \ , \\
\exp\left(-\gamma|\kappa |\Big[1+i\beta \frac{2}{\pi}\mathrm{sgn}(\kappa )\log(\gamma|\kappa |)\Big]+i\mu_0\kappa \right) & \mathrm{for} \quad \alpha=1 \ .  
\end{array}
\right.
\label{eq3.2.2}
\end{equation}
}
The stability parameter $\alpha\in(0, 2]$, strongly affects the tails of the distribution. When $\alpha=2$, the distribution is normal with mean $\mu_0$ and variance $2\gamma^2$, \textit{i.e.}, $\mathcal{N}(\mu_0,2\gamma^2)$. When $0<\alpha<2$, the distribution is non-normal with heavy tails. The shape parameter $\beta\in [-1, 1]$ describes the skewness of the distribution, which is distinguished from the classical skewness defined by $s=\langle (x-\mu)^3\rangle / \sigma^3$, where $\mu$ is the mean of $x$ and $\sigma$ is the standard deviation of $x$. If $\beta=0$, the  distribution is symmetric. If $\beta>0$ ($\beta<0$), the distribution is right (left) skewed. In addition, the scale parameter $\gamma$ is restricted to $\gamma>0$ and the location parameter $\mu_0$ is restricted to $\mu_0\in\mathbb{R}$. The fit parameters of the proper stable distributions are listed in Table~\ref{tab3.2}. The non-normality and the asymmetry of the distributions are revealed by the values of $\alpha$ and $\beta$. Theses stable distributions will be used to work out the probabilities for given cross-responses in Sect.~\ref{sec5.2}.

\subsection{Asymmetry of the market structure}
\label{sec3.3}

Considering the whole market structure, we further quantify the asymmetry along the diagonal of each response matrix. For a $N\times N$ square matrix $X$, the asymmetry of $X$ can be quantified by $\Lambda (X)$,
\begin{equation}
\Lambda (X) = \frac{|| X-X^T ||}{2 || Y ||} \ ,
\label{eq3.3.1}
\end{equation}
where the square matrix $Y$ is defined by $Y_{ij}=X_{ij}(1-\delta_{ij})$ and where $|| X ||$ is the Euclidean norm
\begin{equation}
|| X ||=\sqrt{\sum_{i=1}^{N}\sum_{j=1}^{N}X_{ij}^2} \ .
\label{eq3.3.2}
\end{equation}
The Kronecker delta $\delta_{ij}$ is used to exclude the diagonal elements from $X$. In particular, $\Lambda (X)=0$ means that the matrix $X$ is symmetric along the diagonal, while $\Lambda (X)=1$ indicates that $X$ is anti-symmetric along the diagonal.  A value of $0<\Lambda (X) <1$ arises, when $X$ is asymmetric, where large values of $\Lambda (X)$ indicate high asymmetry. To stabilize the measurement, we introduce the following averaging procedure. Let $k$ be an integer with $1\leq k \leq N$ and let
\begin{equation}
\Xi^{(k|n)}= \left[\begin{array}{ccc}X_{nn} & \cdots & X_{n(n+k-1)} \\ \vdots & \ddots& \vdots \\X_{(n+k-1)n}  & \cdots & X_{(n+k-1)(n+k-1)}\end{array}\right] 
\label{eq3.3.3}
\end{equation}
be a $k\times k$ sub-matrix over the diagonal constructed from $X$, with $1\leq n\leq N-k+1$, then 
\begin{equation}
\langle\Lambda (\Xi^{(k|n)})\rangle=\frac{1}{N-k+1}\sum_{n=1}^{N-k+1} \Lambda \left ( \Xi^{(k|n)} \right)\ .
\label{eq3.3.4}
\end{equation} 
is the average asymmetry of all $k\times k$ sub-matrices in $X$. At a fixed dimension $k$, the averaging over the index $n$ in Eq.~\eqref{eq3.3.4} rules out the influence of elements on the structural asymmetry of $X$. By a further average 
\begin{equation}
\langle\Lambda (X)\rangle=\frac{1}{N}\sum_{k=1}^{N}\langle\Lambda (\Xi^{(k|n)})\rangle \ ,
\label{eq3.3.5}
\end{equation}
we obtain the stabilized overall asymmetry of $X$, as the averaging over the index $k$ in Eq.~\eqref{eq3.3.5} eliminate the influence of dimensions on the structural asymmetry of $X$. Importantly, the diagonal of the sub-matrices $\Xi^{(k|n)}$ lies on the diagonal of $X$. Thus, each $X_{ij}$ is compared with the corresponding $X_{ji}$.

\begin{figure*}[tb]
    \begin{center}
    \includegraphics[width=0.85\textwidth]{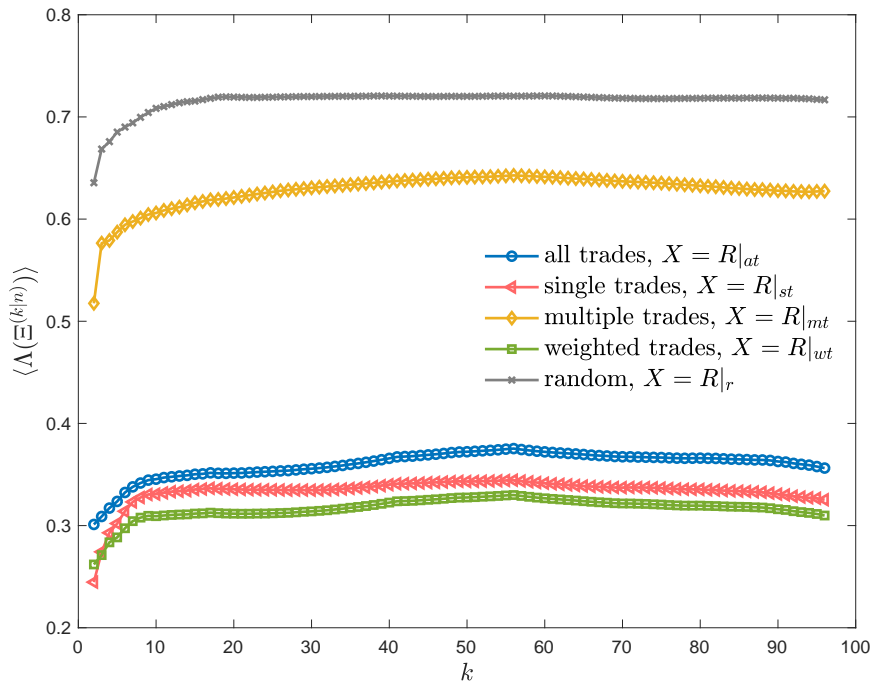}
   \caption{The asymmetries of response matrices $R$ for four cases and of random matrix $R|_\mathrm{r}$ versus the dimension $k$ of matrices.}
   \label{fig3.3}
   \end{center}
\end{figure*}

For our study, the matrix $X$ is the response matrix $R$. The average asymmetry of $R$ versus the matrix dimension $k$ is shown in Fig.~\ref{fig3.3}. We find that the asymmetry for a given response matrix is close to a constant, independent of the dimension $k$, provided $k$ is larger than 10 or so. The values of the stabilized overall asymmetries $\langle\Lambda (R)\rangle$ are listed in Table~\ref{tab3.1}. In all cases, symmetry of the response matrices is absent. Hence, the price impact from stock $j$ to stock $i$ is unequal to the one from stock $i$ to stock $j$, \textit{i.e.}, $\mathcal{G}_{ij}\neq \mathcal{G}_{ji}$ in the model setting. The non-equivalence, for example, in the case of single trades has a deviation of $33.4\%$, hinting at a possibility for arbitrage if ignoring the bid-ask spread. The strong asymmetry in the response structure coincides with the finding that the empirical cross-impact violates the symmetry condition of ``no dynamic arbitrage''~\cite{Schneider2018}.

\section{Eigenvalue spectra of asymmetric market structures}
\label{sec4}

Random matrices have been used to analyze the spectrum properties of cross-correlations of financial data~\cite{Laloux1999,Laloux2000,Plerou2000,Plerou2002,Potters2005,Bouchaud2009}. The eigenvalue spectrum of a random correlation matrix is well known to be given by the Mar\u{c}enko-Pastur distribution~\cite{Marvcenko1967,Silverstein1995,Sengupta1999}. These random correlation matrices are symmetric and the corresponding eigenvalues are real. However, for an asymmetric random matrix, the eigenvalues are complex and a description with the Mar\u{c}enko-Pastur distribution is not appropriate. To analyze the eigenvalue spectrum of an asymmetric matrix $X$, we decompose the matrix into a symmetric part $X_{S}$ and an asymmetric part $X_{A}$,
\begin{equation}
X=X_{S}+X_{A} \ ,
\label{eq4.1}
\end{equation}
with
\begin{equation}
X_{S}=(X+X^T)/2  
\qquad \mathrm{and} \qquad 
X_{A}=(X-X^T)/2 \ . 
 \label{eq4.2} 
\end{equation}
Thus, the asymmetry of $X$ is fully accounted for by $X_{A}$. As $X_{A}$ is anti-symmetric, the non-zero eigenvalues $\lambda_k$ of $X_{A}$ are purely imaginary, given by
\begin{equation}
X_{A}\psi_k=\lambda_k \psi_k \ ,
\label{eq4.3}
\end{equation}
where $\psi_k$ is the corresponding eigenvector.

The distribution of eigenvalues of random asymmetric matrices was computed by Sommers \textit{et al.}~\cite{Sommers1988}. For an ensemble of $N\times N$ random asymmetric matrices $M$, where the elements $M_{ij}$ are normally distributed with zero mean and correlations
\begin{equation}
\langle M_{ij}^2 \rangle=1
\qquad \mathrm{and} \qquad
\langle M_{ij}M_{ji} \rangle=c \ 
\label{eq4.4}
\end{equation}
for $i\neq j$ and $-1\leq c \leq 1$,
the average density $p(\omega_k)$ of eigenvalues $\omega_k=x_k+iy_k$ is given by 
\begin{eqnarray}
p(\omega_k)=\left \{ 
\begin{array}{ll}
(\pi ab)^{-1}\ , & \mathrm{if~} (x_k/a)^2+(y_k/b)^2\leq 1 \ , \\
0\ , & \mathrm{otherwise} \ ,
\end{array}
\right.
\label{eq4.5}
\end{eqnarray} 
where $a=1+c$ and $b=1-c$. The cases $c=1$ and $c=0$ correspond to ensembles of symmetric matrices and fully asymmetric matrices in which $M_{ij}$ and $M_{ji}$ are independent, respectively. When $c=-1$, the matrix is anti-symmetric, \textit{i.e.}, $M_{ij}=-M_{ji}$, with non-zero imaginary eigenvalues $\pm iy_k$. The projection of $p(\omega_k)$ on the imaginary axis leads to a generalized semicircle law, which describes the probability density distribution
\begin{equation}
p(y_k)=\int dx_k p(\omega_k)=\frac{2}{\pi b^2} (b^2-y_k^2)^{1/2} \ , \quad |y_k|\leq b \ .
\label{eq4.6}
\end{equation}
At the points $y_k=\pm b$, the probability densities are equal to zero. 

\begin{figure*}[tbp]
\begin{center}
    \includegraphics[width=1\textwidth]{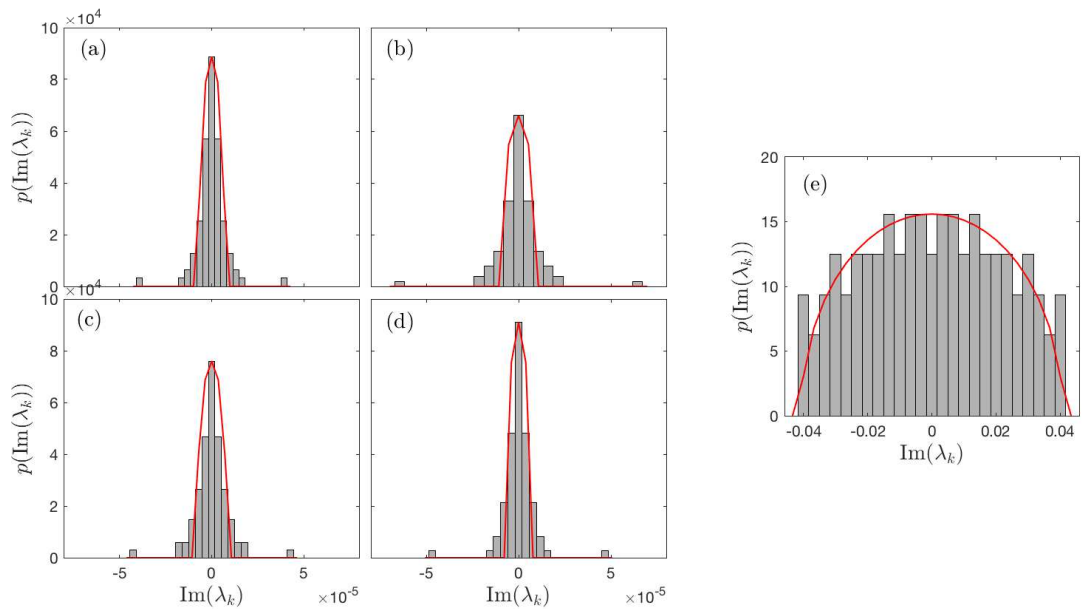}
   \caption{Probability density distributions of imaginary parts of eigenvalues, (a) case of all trades, (b) case of single trades, (c) case of multiple trades, (d) case of weighted trades, (e) random case. For comparison, the distributions~\eqref{eq4.6} are displayed as red lines.}
   \label{fig4.1}
\end{center}
\end{figure*}

Using Eq.~\eqref{eq4.2}, we calculate for all response matrices $R$ the corresponding asymmetric matrices $R_{A}$. We compute the eigenvalues of $R_{A}$ and work out the probability density distributions $p(\mathrm{Im}(\lambda_{k}))$, shown in Fig.~\ref{fig4.1}. The histograms in Fig.~\ref{fig4.1} are normalized to one. All these distributions are compared with the distributions of random matrices drawn from Eq.~\eqref{eq4.6}. Here, due to the anti-symmetry of $R_A$, $c=-1$, such that $b=2$. However, the probability densities of $\mathrm{Im}(\lambda_{k})$ need to be rescaled to match the normalized histograms. A convenient way is to scale $b$ by a factor of $(\pi p(0))^{-1}$, because for a normalized histogram, Eq.~\eqref{eq4.6} at the point $y_k=0$ results in $b=2/(\pi p(0))$.

As expected, the distribution resulting from Eq.~\eqref{eq4.6} matches well the distribution for the random case. In contrast, it deviates largely from each of the distributions for the four types of responses. The differences are significant at the tails of distributions where the largest imaginary parts of eigenvalues are located. In the previous studies~\cite{Plerou2002,Suparno2016}, the largest eigenvalue for cross-correlation matrices are interpreted as the market mode or collective ``response" of the whole market to stimuli, such as economic growth, interest rate increase, or political events. Because the eigenvector corresponding to the largest eigenvalue has nonzero components everywhere, the influence represented by the largest eigenvalue is common to all stocks. In our case, the eigenvectors corresponding to the largest imaginary parts of eigenvalues are complex. Besides the difficulty from processing the complex eigenvectors, a lack of relevant data makes it impossible to identify such market or sector mode. However, the non-random distributions of eigenvalue spectra suggest that the response matrices contain asymmetric information. If the market is efficient, the prices should perfectly reflect the information that is available~\cite{Fama1970}. As a result, the immediate responses between assets are symmetric~\cite{Schneider2018}, \textit{i.e.}, $R_{ij}=R_{ji}$ and the arbitrage opportunities are absent in the market. However, the asymmetric information we find reveals that the market is not fully efficient and arbitrage opportunities can arise. Without such opportunities, there will be no incentives for traders to acquire information and the price discovery aspect of financial markets will cease to exist~\cite{Grossman1980,Lo2004}. The asymmetric information thus plays a role in maintaining a natural market ecology, where the possible arbitrage opportunities provide the motivation for traders to stay in the market.

\section{Entropy of asymmetric market structures}
\label{sec5}

We introduce the Shannon entropy~\cite{Shannon1948} to quantify the randomness of eigenvalue spectra in Sect.~\ref{sec5.1}. Making use of the stable distributions, we map the response matrices into probability matrices and compute the entropy of price impacts in Sect.~\ref{sec5.2}. For given entropy matrices, we construct directional networks for impacts and further explore how the network evolves with the entropy in Sect.~\ref{sec5.3}. Since the price impact for a single trade is estimated by the immediate response, we use the response for convenience to represent the price impact in the following.

\subsection{Entropy of eigenvalue spectra}
\label{sec5.1}

We used the eigenvalue spectrum to identify the non-randomness of asymmetric price impacts, but we have not yet quantified the randomness. To this end, we resort to the entropy in information theory, also known as the Shannon entropy~\cite{Shannon1948}, which is capable of analyzing the randomness and unpredictability of a system. The Shannon entropy is defined as
\begin{equation}
H(Z)=-\sum_{k=1}^{n}P(z_k)\log_{\xi}P(z_k) \ .
\label{eq5.1.1}
\end{equation}
Here, $Z$ is a discrete random variable with possible values $\{z_1,\cdots,z_n\}$, and $P(z_k)$ is the probability of the value $z_k$. We notice that $P(z_k)$ is, in a discrete setting, a probability, not a probability density. The base of the logarithm used is $\xi$. The Shannon entropy measures the average amount of information in the data. If the entropy is high, the amount of information that can be measured is small because much useful information is hidden in the random noise. Let the randomness estimate the strength of random noise in a system. To hide the useful information, the randomness of the system must be large.

To quantify the randomness of the eigenvalue spectrum, we replace $z_k$ with $\mathrm{Im} (\lambda_k)$ and set the base $\xi$ to Euler's number $e$. We notice that a probability $P(\mathrm{Im} (\lambda_k))$ of zero leads to a zero contribution in Eq.~\eqref{eq5.1.1}, as $P(\mathrm{Im} (\lambda_k))\cdot\log_{\xi}P(\mathrm{Im} (\lambda_k))$ also vanishes in that case. The resulting entropies of eigenvalue spectra for the four types of responses and the random case are listed in Table~\ref{tab3.1}. Among the four types of responses, the case of weighted trades has the lowest entropy, implying a larger amount of private information. The case of single trades is second to the case of weighted trades. For comparison, the random case presents the highest entropy and obviously lacks useful information.

\subsection{Entropy of price impacts}
\label{sec5.2}

With the probabilities of responses for pairs of stocks, we can measure the entropy of impacts for the whole market. Using the stable distributions which we worked out in Sect.~\ref{sec3}, we calculate the probability $P_0(R_{ij})$ for a given value of cross-responses $R_{ij}$ by
\begin{equation}
P_0(R_{ij})=P_0[E_{k}\leq R_{ij}< E_{k+1}]=\sum_{x=E_{k}}^{E_{k+1}}p(x)\Delta x  \ ,
\label{eq5.2.1}
\end{equation}
where $[E_{k}, E_{k+1})$ is the interval containing $R_{ij}$. The index $k$ takes values $1,\cdots,K$ if the data is grouped in $K$ bins. The last bin also includes the right-most bin edge, \textit{i.e.}, $E_{K}\leq R_{ij}\leq E_{K+1}$. The width of the bins, \textit{i.e.}, $E_{k+1}-E_{k}$, is the same for all bins. Hence, for a fixed pair ($i$, $j$), we have $\sum_{R_{ij}}P_0(R_{ij})=1$ by summing over all discrete values of $R_{ij}$. We then map the response matrix to a probability matrix, with entries 
\begin{equation}
P(R_{ij})=\frac{P_0(R_{ij})}{\sum\limits_{k=1}^{N}\sum\limits_{l=1,l\neq k}^{N} P_0(R_{kl})} \ . 
\label{eq5.2.2}
\end{equation}   
Here, we ignore the case of self-responses by setting $P(R_{ii})\\ =1$ for the probability of the self-response, such that the contribution $P(R_{ii})\log_{\xi}P(R_{ii})=0$ vanishes. Equation~\eqref{eq5.2.2} defines a probability for all discrete values $R_{ij}$, running over all pairs ($k$, $l$), where $k\neq l$.

\begin{figure*}[tbp]
\begin{center}
    \includegraphics[width=0.8\textwidth]{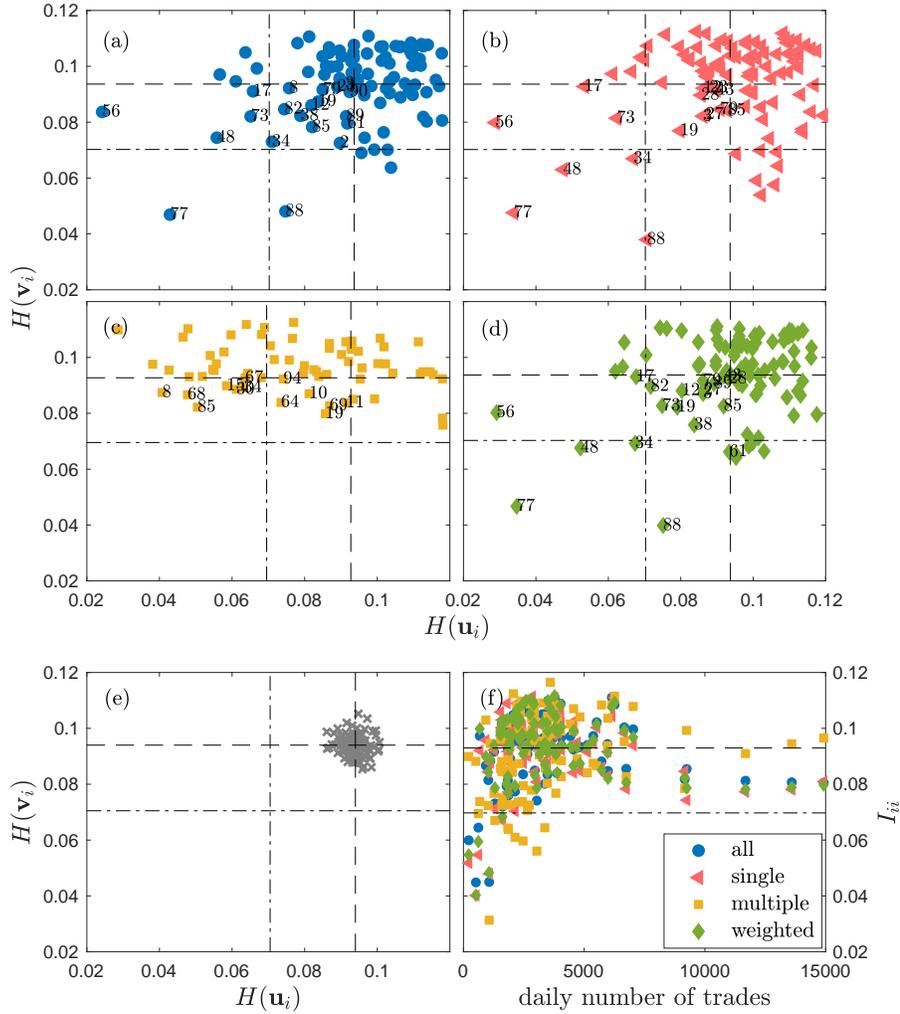}
   \caption{Scatter plots of stock $i$ located at the position ($H(\vec{u}_i)$, $H(\vec{v}_i)$) with (a) case of all trades, (b) case of single trades, (c) case of multiple trades, (d) case of weighted trades, and (e) random case. (f) The entropy of impacts $I_{ii}(R)$ of stock $i$ for the four types of responses versus the daily number of trades of the stock itself. The dash lines indicate the positions of $\langle H(\vec{u}_i)\rangle$ and $\langle H(\vec{v}_i)\rangle$ in (a)--(e) and the position of $\langle I_{ii}\rangle$ in (f); the dot-dash lines indicate the positions of $0.75\langle H(\vec{u}_i)\rangle$ and $0.75\langle H(\vec{v}_i)\rangle$ in (a)--(e) and the position of $0.75\langle I_{ii}\rangle$ in (f). The number near each mark is an index of a stock listed in~\ref{appA}.}
   \label{fig5.1}
  \end{center}
\end{figure*}

\begin{figure*}[htbp]
\begin{center}
    \includegraphics[width=0.95\textwidth]{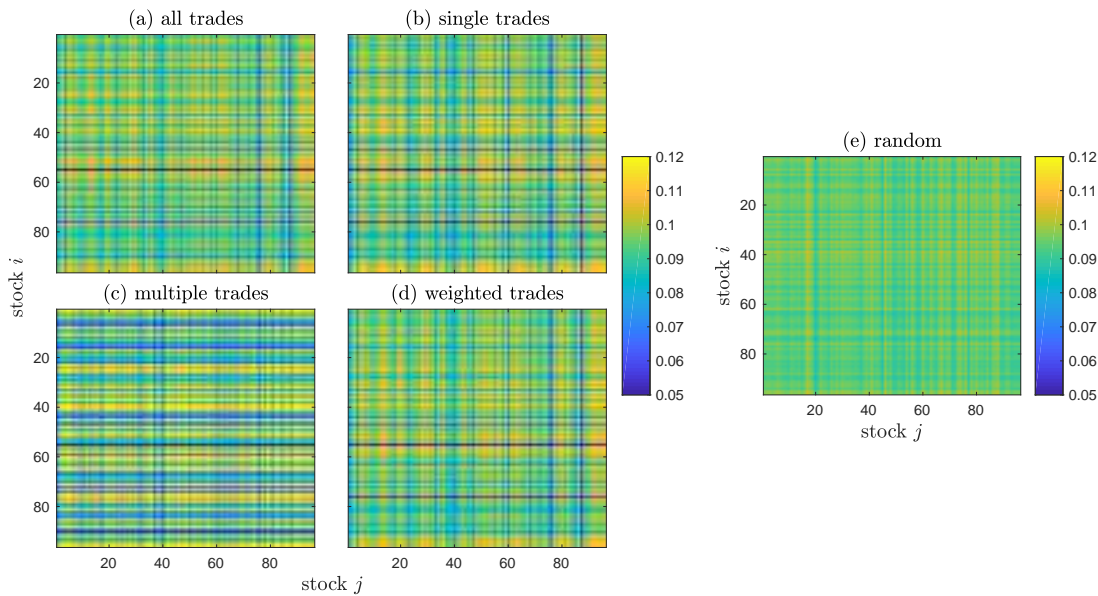}
   \caption{Entropy matrices of impacts $I$ in (a) case of all trades, (b) case of single trades, (c) case of multiple trades, (d) case of weighted trades, and (e) random case. }
   \label{fig5.2}
\end{center}
\end{figure*}

To measure the randomness of the impacts for given stocks, we write the response matrix in terms of its rows $\vec{u}_i^T$, $i=1,\cdots, N$, and columns $\vec{v}_j$, $j=1,\cdots, N$, where
\begin{equation}
R=\left[\begin{array}{c}\vec{u}_1^T \\ \vec{u}_2^T \\ \vdots \\ \vec{u}_N^T\end{array}\right]
\qquad \mathrm{and} \qquad
R=[\vec{v}_1 \ \vec{v}_2 \ \cdots\ \vec{v}_N] \ .
\label{eq5.2.3}
\end{equation}
Here, $\vec{u}_i$ captures the target reaction of the impact, \textit{i.e.}, the price change, and $\vec{v}_j$ the triggering effect, \textit{i.e.}, the trading information. Their randomness can be quantified by the entropies of the row vector $\vec{u}_i$ and of the column vector $\vec{v}_j$, respectively,
\begin{eqnarray}
H(\vec{u}_i)=-\sum_{j=1}^{N}P(R_{ij})\log_{\xi}P(R_{ij}) \ ,  \label{eq5.2.4} \\
H(\vec{v}_j)=-\sum_{i=1}^{N}P(R_{ij})\log_{\xi}P(R_{ij}) \ .  \label{eq5.2.5} 
\end{eqnarray}
In the following, we use the natural logarithm, $\xi=e$. The randomness of the price changes as well as of the trading information affect the price impact. Therefore, we define the entropy of impacts as
\begin{equation} 
I_{ij}=[H(\vec{u}_i) H(\vec{v}_j)]^{1/2}
\label{eq5.2.6}
\end{equation}
to weight the randomness of the impact between stocks $i$ and $j$. Large entropy indicates high randomness, and small one low randomness of the impacts.

In Fig.~\ref{fig5.1}, we display scatter plots of stock $i$ located at the position ($H(\vec{u}_i)$, $H(\vec{v}_i)$) in the entropy plane. As seen, the four cases differ strongly from the random one, in which all points scatter isotropically around the mean values ($\langle H(\vec{u}_i)\rangle$, $\langle H(\vec{v}_i)\rangle$). In contrast, in the four cases a small fraction of stocks falls into the regions $0<H(\vec{u}_i)\leq \langle H(\vec{u}_i)\rangle$ and $0<H(\vec{v}_i)\leq \langle H(\vec{v}_i)\rangle$, and the smaller regions $0<H(\vec{u}_i)\leq 0.75\langle H(\vec{u}_i)\rangle$ and $0<H(\vec{v}_i)\leq 0.75\langle H\\ (\vec{v}_i)\rangle$. For these stocks, both the triggering effects and target reactions of impacts are less affected by the random noise. Accordingly, the private information encoded in the corresponding prices is more likely to be extracted. Outstanding are the stocks with indices 77, 88, 56 and 48, which have small daily number of trades. Among their countable trades, a trade is noticeable and the triggering effect for an impact, \textit{i.e.}, its trading information, is easily spread without much interference with the random noise. When these less traded stocks have less liquidity as well, due to the large bid-ask spread, the target reactions of impacts, \textit{i.e.}, their price changes, become pronounced. We thus find a relation between the average daily number of trades and the entropy of impacts. As shown in Fig.~\ref{fig5.1} (f), the least traded stocks have the lowest entropy of impacts, which is smaller than $0.75\langle I_{ii}\rangle$, while the most frequently traded stocks have an entropy between $0.75\langle I_{ii}\rangle$ and $\langle I_{ii}\rangle$. Assuming that values larger than $\langle I_{ii}\rangle$ indicate presence of randomness, most of the stocks with an average number of trades show randomness either for the trading information or for the price change, or for both.

\subsection{Networks of price impacts}
\label{sec5.3}

\begin{figure*}[tbp]
\begin{center}
    \includegraphics[width=0.92\textwidth]{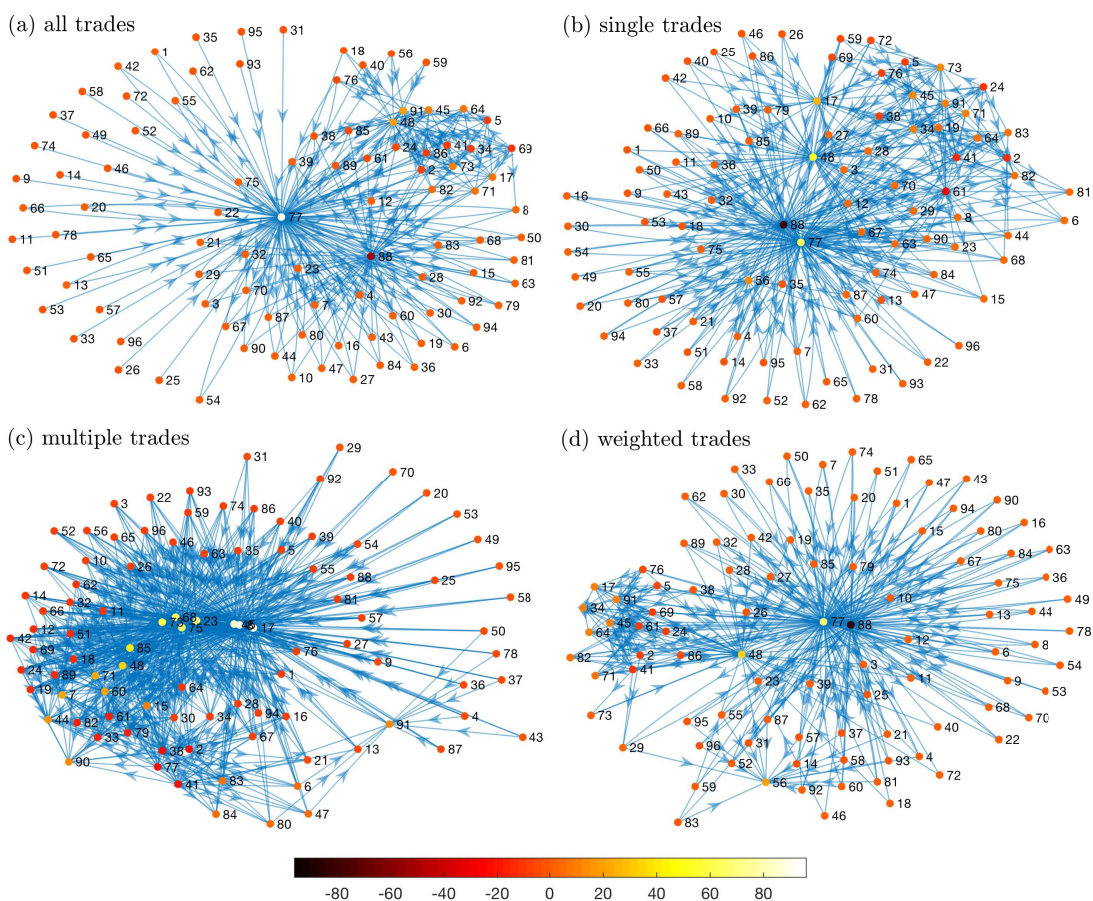}
   \caption{Networks of impacts with the entropy $0.6\langle I_{ij}\rangle<I_{ij}\leq 0.75\langle I_{ij}\rangle$ in (a) case of all trades, (b) case of single trades, (c) case of multiple trades and (d) case of weighted trades. The start of an arrow $j\rightarrow i$ represents the impacting stock $j$ with trading information and the end of the arrow represents the impacted stock $i$ with price changes. The colour of nodes indicates the connectivity of nodes. A positive value of the connectivity is equal to the ingoing connectivity when it is larger than the outgoing connectivity, while a negative value of the connectivity is equal to $-1$ multiplied by the outgoing connectivity when it is larger than the ingoing connectivity.}  
   \label{fig5.3}
\end{center}
\end{figure*}

To further characterize the randomness of impacts across different stocks, we introduce entropy matrices $I$ with entries $I_{ij}$, shown in Fig.~\ref{fig5.2}, where the case $i=j$ is also included. The structures in the four types of responses clearly visualize the degrees of information and randomness, whereas the picture is blurred in the random case. To quantify the impact among stocks, we define the distance between two stocks as the entropy $I_{ij}$ in a given range and as zero if $I_{ij}$ is out of that range. Thereby, we are able to construct a directional network of impacts. For instance, in the range $0.6\langle I_{ij}\rangle<I_{ij}\leq 0.75\langle I_{ij}\rangle$, the networks of impacts for the four non-random cases are shown in Fig.~\ref{fig5.3}. The centering stocks, such as the ones indexed by 77 and 88, have the highest in- and outgoing connectivities. Here, the ingoing (outgoing) connectivity is the number of edges connected to the node according to ingoing (outgoing) direction of the arrows. 

\begin{figure*}[htbp]
\begin{center}
    \includegraphics[width=0.92\textwidth]{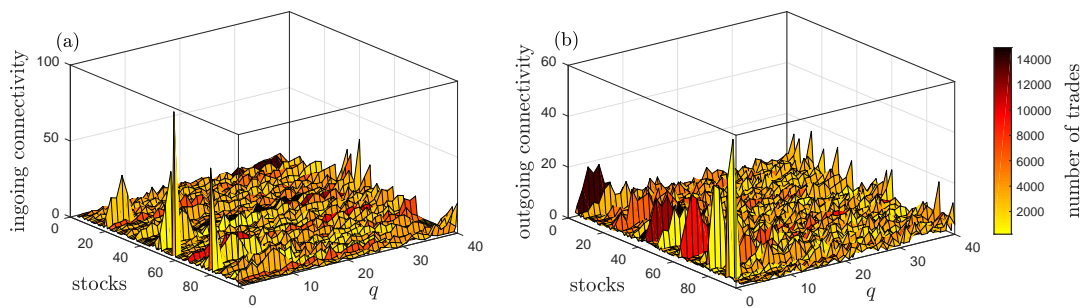}
   \caption{(a) Ingoing connectivity and (b) outgoing connectivity of each stock depending on the group $q$. With the increase of $q$, the entropy of impacts increases and the randomness increases. The colour indicates the average daily number of trades of each stock.}
   \label{fig5.4}
\end{center}
\end{figure*}

We take a closer look at how the network evolves with the entropy of impacts. To this end, we only consider the case of single trades and process the data as follows: 
\begin{enumerate}
\item[(1)] we rank in total 9120 values of $I_{ij}$ with $i\neq j$ from the $96\times 96$ entropy matrix $I$ in ascending order;
\item[(2)] we form groups of 228 values of $I_{ij}$ each in ascending order and label each group by $q$ with $q=1,2,\cdots,40$, such that with increase of $q$, the entropy of impacts increases;
\item[(3)] we extract the entropy matrix $I^{(q)}$ for the $q$-th group with entries $I_{ij}^{(q)}$ defined by
\begin{equation}
I_{ij}^{(q)}=\left\{
\begin{array}{ll}
I_{ij} \ , &\quad \textrm{if $I_{ij}$ is in the $q$-th group} \ , \\
0	\ , & \quad \mathrm{otherwise} \ . 
\end{array}
\right.
\label{fig5.3.1}
\end{equation}
\end{enumerate}
With the entropy matrix $I^{(q)}$, we construct the impact network for the $q$-th group. The network is characterized by the in- and outgoing connectivities of each stock. The dependencies of in- and outgoing connectivities on the stocks and groups are shown in Fig.~\ref{fig5.4}, where the colour denotes the average daily number of trades of each stock. Remarkably, the in- and outgoing connectivities are larger when the stocks are present in the groups with small entropy of impacts. Increasing the entropy of impacts makes the network random with little structure. This demonstrates that the impact with small entropy indeed reveals useful information. On the other hand, the stocks with a higher ingoing connectivity have a smaller average daily number of trades, but the stocks with a higher outgoing connectivity do not feature small number of trades. In particular, the stocks with a large daily number of trades, \textit{e.g.}, AAPL indexed by 2, FB indexed by 38, GILD indexed by 41, and MSFT indexed by 61, are more likely to impact other stocks. The result that impacts are related to the number of trades is consistent with the findings in Ref.~\cite{Wang2016b}, where another method of analysis was used.

\section{Conclusions} 
\label{sec6}
 
By reconstructing the order book, we worked out the price cross-response, \textit{i.e.}, the price change of one stock due to single trades or multiple trades of another stock. As we focused on the immediate response, we used an event time scale. The cross-responses are averaged over all trades, as well as over single trades, multiple trades and trades weighted by the percentages of single trades and multiple trades. Distinguishing these four types of cross-responses yields a detailed picture of the price impacts. 

The distributions of cross-responses for the whole market show right skewness, revealing the imbalance of positive and negative cross-responses. This implies that a buy (or a sell) of one stock is very likely to move up (or down) the price of another stock. The distributions are fitted well by stable distributions. The fit parameters reflect the asymmetry and may be interpreted as measuring the degree of non-randomness of the events. By quantifying the asymmetry for the cross-responses, we found that the impact from stock $j$ to $i$ is not equal to the one from stock $i$ to $j$. This corroborates the findings in Ref.~\cite{Schneider2018} and hints at a possibility for arbitrage when ignoring the bid-ask spread.  

We also evaluated eigenvalue spectra of asymmetric impact structures. The results demonstrate that the information encoded in the asymmetric impact structures is not fully random. The Shannon entropy~\cite{Shannon1948} reveals that the cases of single trades and of weighted trades contain more non-random information than others. We further estimated the entropy of impacts, which is composed of the entropy of trading information and of price changes. For a given entropy of impacts, we constructed a directional network to visualize the impacts among stocks. The evolution of this network discloses that impacts with small entropy are more informative. Furthermore, the stocks with large daily numbers of trades are more likely to impact others while the less frequently traded stocks tend to be impacted by others. 

We identified, quantified and visualized the asymmetric information in price impacts and found that (1) the impacts in the whole market are asymmetric and non-random; (2) the randomness of impacts across stocks can be quantified by the entropy of impacts; (3) informative impacts are present at small entropy; (4) the stocks with large (small) number of trades are likely to affect (be affected by) others.

\section*{Acknowledgments}
\addcontentsline{toc}{section}{Acknowledgments}
We thank A. Becker, S. Krause, Y. Stepanov and D. Waltner for fruitful discussions.

\begin{appendices}
\setcounter{figure}{0}
\setcounter{equation}{0}
\setcounter{table}{0}
\renewcommand{\thetable}{\Alph{section}.\arabic{table}}

\section{Trading data}
\label{appA}

By order reconstruction, we filter the trade and quote data to a resolution of one millisecond. Table~\ref{appA.1} lists the daily numbers of trades and quotes and the spread between the best ask and bid for 96 stocks in NASDAQ 100 index. The data is measured during intraday trading time from 9:40 to 15:50 EST and averaged over five trading days from March 7th to March 11th in 2016.

\begin{table*}[htbp]
\newcolumntype{C}[1]{>{\centering}m{#1}}
\caption{Daily trade and quote information averaged over five trading days}
\begin{center}
\begin{footnotesize}
\begin{tabular}{|C{0.3cm}C{1.2cm}rrC{1.3cm}|C{0.3cm}C{1.2cm}rrc|} 
\hline
No. &\makecell{Stock \\symbol}&\makecell{Number \\of trades} &\makecell{Number \\of quotes} &\makecell{Spread \\(dollars)} &  No. &\makecell{Stock \\symbol}&\makecell{Number \\of trades} &\makecell{Number \\of quotes} &\makecell{Spread \\(dollars)}  \\
\hline
1	&	AAL		&   4,563   &   86,481   &   0.013   &   	49	&	JD		&   3,596   &   64,622   &   0.012   \\   
2	&	AAPL	&   13,598   &   312,068   &   0.012   &   	50	&	KHC		&   3,140   &   45,201   &   0.027   \\   
3	&	ADBE	&   4,553   &   41,067   &   0.034   &   	51	&	KLAC	&   1,803   &   53,830   &   0.014   \\   
4	&	ADI		&   2,931   &   30,598   &   0.023   &   	52	&	LBTYA	&   2,759   &   60,328   &   0.013   \\   
5	&	ADP		&   2,954   &   59,388   &   0.023   &   	53	&	LLTC	&   2,300   &   59,533   &   0.014   \\   
6	&	ADSK	&   3,389   &   33,981   &   0.032   &   	54	&	LMCA	&   1,585   &   30,686   &   0.016   \\   
7	&	AKAM	&   2,439   &   37,125   &   0.024   &   	55	&	LRCX	&   3,826   &   40,132   &   0.038   \\   
8	&	ALXN	&   2,466   &   25,072   &   0.212   &   	56	&	LVNTA	&   1,063   &   10,127   &   0.046   \\   
9	&	AMAT	&   2,066   &   61,402   &   0.010   &   	57	&	MAR		&   3,495   &   54,363   &   0.024   \\   
10	&	AMGN	&   5,132   &   31,036   &   0.083   &   	58	&	MAT		&   2,918   &   50,904   &   0.012   \\   
11	&	AMZN	&   5,376   &   54,185   &   0.428   &   	59	&	MDLZ	&   3,666   &   74,011   &   0.011   \\   
12	&	ATVI		&   3,882   &   110,135   &   0.011   &   	60	&	MNST	&   1,591   &   12,606   &   0.123   \\   
13	&	AVGO	&   5,518   &   39,476   &   0.095   &   	61	&	MSFT	&   9,245   &   164,847   &   0.011   \\     
14	&	BBBY	&   2,590   &   29,066   &   0.026   &   	62	&	MU		&   2,351   &   62,667   &   0.010   \\   
15	&	BIDU	&   2,729   &   18,414   &   0.191   &   	63	&	MYL		&   5,969   &   56,840   &   0.017   \\   
16	&	BIIB		&   2,818   &   21,057   &   0.273   &   	64	&	NFLX	&   9,164   &   73,919   &   0.049   \\   
17	&	BMRN	&   2,135   &   17,778   &   0.180   &   	65	&	NTAP	&   2,210   &   54,022   &   0.012   \\   
18	&	CA		&   1,531   &   37,172   &   0.011   &   	66	&	NVDA	&   2,935   &   78,995   &   0.011   \\   
19	&	CELG	&   6,742   &   45,181   &   0.056   &   	67	&	NXPI		&   3,824   &   34,424   &   0.056   \\   
20	&	CERN	&   3,440   &   34,198   &   0.022   &   	68	&	ORLY	&   1,837   &   21,748   &   0.349   \\   
21	&	CHKP	&   2,030   &   15,733   &   0.048   &   	69	&	PAYX	&   1,838   &   41,534   &   0.013   \\   
22	&	CHRW	&   2,021   &   19,541   &   0.032   &   	70	&	PCAR	&   3,315   &   38,421   &   0.019   \\   
23	&	CHTR	&   2,650   &   20,287   &   0.189   &   	71	&	PCLN	&   1,029   &   17,369   &   2.525   \\   
24	&	CMCSA	&   5,984   &   179,820   &   0.011   &   	72	&	QCOM	&   7,030   &   117,669   &   0.011   \\   
25	&	COST	&   3,487   &   21,260   &   0.059   &   	73	&	REGN	&   1,300   &   13,691   &   0.883   \\   
26	&	CSCO	&   3,273   &   106,513   &   0.011   &   	74	&	ROST	&   3,868   &   61,006   &   0.016   \\   
27	&	CTSH	&   4,823   &   54,413   &   0.016   &   	75	&	SBAC	&   1,935   &   19,128   &   0.084   \\   
28	&	CTXS	&   2,477   &   22,161   &   0.053   &   	76	&	SBUX	&   5,719   &   111,888   &   0.012   \\   
29	&	DISCA	&   3,152   &   72,949   &   0.012   &   	77	&	SIRI		&   514   &   310,522   &   0.010   \\   
30	&	DISH	&   2,261   &   27,956   &   0.026   &   	78	&	SNDK	&   3,687   &   43,642   &   0.021   \\   
31	&	DLTR	&   4,021   &   26,372   &   0.037   &   	79	&	SPLS	&   1,108   &   37,419   &   0.010   \\   
32	&	EA		&   4,708   &   53,137   &   0.021   &   	80	&	SRCL	&   1,588   &   11,546   &   0.094   \\   
33	&	EBAY	&   2,850   &   69,949   &   0.011   &   	81	&	STX		&   4,056   &   39,986   &   0.017   \\   
34	&	EQIX	&   1,615   &   32,672   &   0.408   &   	82	&	SYMC	&   1,784   &   41,076   &   0.010   \\   
35	&	ESRX	&   6,144   &   64,200   &   0.020   &   	83	&	TRIP		&   3,473   &   23,576   &   0.050   \\   
36	&	EXPD	&   2,310   &   32,064   &   0.015   &   	84	&	TSCO	&   1,535   &   15,998   &   0.073   \\   
37	&	FAST	&   2,816   &   32,818   &   0.018   &   	85	&	TSLA	&   3,367   &   19,876   &   0.277   \\   
38	&	FB		&   14,921   &   213,281   &   0.016   &   	86	&	TXN		&   3,479   &   81,678   &   0.012   \\   
39	&	FISV		&   1,856   &   22,382   &   0.043   &   	87	&	VIAB		&   3,769   &   35,708   &   0.023   \\   
40	&	FOXA	&   2,388   &   66,595   &   0.010   &   	88	&	VIP		&   217   &   6,494   &   0.010   \\   
41	&	GILD		&   11,681   &   95,291   &   0.016   &   	89	&	VOD		&   926   &   62,212   &   0.011   \\   
42	&	GOOG	&   4,426   &   40,123   &   0.515   &   	90	&	VRSK	&   1,264   &   11,367   &   0.044   \\   
43	&	GRMN	&   1,909   &   30,404   &   0.017   &   	91	&	VRTX	&   3,037   &   19,411   &   0.129   \\   
44	&	HSIC	&   674   &   13,076   &   0.226   &   		92	&	WDC	&   6,662   &   61,126   &   0.026   \\   
45	&	ILMN	&   1,860   &   16,979   &   0.266   &   	93	&	WFM	&   3,775   &   63,946   &   0.012   \\   
46	&	INTC		&   3,933   &   115,962   &   0.011   &   	94	&	WYNN	&   4,046   &   26,083   &   0.081   \\   
47	&	INTU		&   2,299   &   23,210   &   0.053   &   	95	&	XLNX	&   2,450   &   33,769   &   0.016   \\   
48	&	ISRG	&   616   &   13,015   &   1.269   &   		96	&	YHOO	&   6,258   &   134,390   &   0.011   \\   
\hline
\end{tabular}
\end{footnotesize}
\end{center}
\label{appA.1}
\end{table*}

\end{appendices}

\end{document}